\documentclass[pre,twocolumn,showpacs,showkeys,nofootinbib,preprintnumbers,
amsmath,amssymb,floatfix]{revtex4-1}

\usepackage{graphicx}
\usepackage{dcolumn}
\usepackage{bm}
\usepackage{epsfig,wrapfig,boxedminipage,enumerate}
\usepackage{nicefrac}
\usepackage{amsmath}
\usepackage{relsize}
\usepackage{epsfig}
\usepackage{amssymb}
\usepackage{color}

\newcommand{\vect}[1]{\mathbf{#1}}

\newcommand{\kt}{k_{\rm B}T}
\newcommand{\one}{\mathcal{I}}

\newcommand{\bea}{\begin{equation}}
\newcommand{\eea}{\end{equation}}
\newcommand{\lG}{l_{\rm G}}

\def\br{\mathbf{r}}
\def\bx{\mathbf{x}}
\def\bk{\mathbf{k}}

\def\bu{\mathbf{u}}
\def\bv{\mathbf{v}}

\def\be{\mathbf{e}}
\def\bj{\mathbf{j}}
\def\bet{\boldsymbol{\eta}}
\def\bom{\boldsymbol{\omega}}
\def\sigh{\sigma_{\rm H}}

\def\lhyd{L_{\rm hydro}}
\def\confell{\ell_\textrm{c}}

\newcommand{\eq}[1]{Eq.~(\ref{#1})}
\newcommand{\eqs}[1]{Eqs.~(\ref{#1})}
\newcommand{\vare}{\varepsilon}

\begin{document}

\title{Onset of anomalous diffusion in colloids confined to
  quasi--monolayers}

\author{J.~Bleibel$^{1,2}$, Alvaro Dom\'\i nguez$^3$, M.~Oettel$^1$} 
\affiliation{$^1$Institut f\"ur angewandte Physik, Universit\"at T\"ubingen,  
Auf der Morgenstelle 10, 72076 T\"ubingen, Germany}
\affiliation{$^2$Max-Planck-Institut f\"ur Intelligente Systeme,
  Heisenbergstr.~3, 70569 Stuttgart, Germany} 
\affiliation{$^3$F\'\i sica Te\'orica, Universidad de Sevilla, Apdo.~1065,
  41080 Sevilla, Spain}

\date{\today}

\begin{abstract}
  It has been recently shown that a colloidal monolayer, e.g., formed
  at a fluid interface or by means of a suitable confining
    potential, exhibits anomalous collective diffusion. This 
  is a consequence of the 
  hydrodynamic interactions mediated by the three--dimensional (3D)
  ambient fluid when the particles are confined to reside on a
  two--dimensional (2D) manifold. We study theoretically and with
  numerical simulations the crossover from normal to anomalous
  diffusion as the particles are, in real systems, confined by a 3D
  external potential and thus have the possibility to fluctuate out of
  the 2D manifold, thus forming actually a quasi--monolayer.
\end{abstract}



\maketitle

\section{Introduction}
\label{sect:intro}
Particle--laden fluid interfaces are a common subject in soft matter physics, 
and offer an interesting approach to effectively 2D
systems for theory and experiment. In many cases, it is a good
simplification to treat the system of fluids, interface and particles
as a genuine 2D problem, e.g., in order to explore phase
transitions~\cite{Zahn:1997,Wille:2002} or clustering behavior in
lower dimensions~\cite{Bleibel:2011}. However, as soon as hydrodynamic
interactions are considered, the full 3D nature of the setup
becomes important~\cite{Bleibel:SMC14}.
A colloidal monolayer is an example of the configuration of
\emph{partial confinement}, as termed in Ref.~\cite{Bleibel:SMC14},
because a part of the components of the system is confined (the
particles are restricted to move in a plane), whereas other
constituents are not confined (the ambient fluid occupies the adjacent
volume).
The dimensional mismatch between the 2D colloidal dynamics and the 3D
hydrodynamic interactions mediated by the ambient fluid flow induces
anomalous diffusion (more precisely, superdiffusion) for the
collective, i.e., large--scale dynamics of the
monolayer~\cite{NKPB02,Bleibel:SMC14,Bleibel:JPCM15,Dom15}.
These theoretical predictions lead to a reinterpretation of
experimental results that had actually measured the 
anomalous collective diffusion~\cite{Lin:2014}.

The dynamics of the spatial distribution of particles in a colloidal
monolayer can be conveniently characterized by a wavenumber dependent
diffusion coefficient $D(k)$, that can be expanded in powers of the
wavenumber,
\begin{equation}
  \label{def_Dk}
  \frac{D(k)}{D_0} = \sum_{n=-\infty}^{\infty} \beta_n k^n,
\end{equation}
where the constant $D_0$ is conventionally taken to be the
single--particle diffusion coefficient in the dilute limit.  Normal
diffusion is then characterized by the absence of negative powers of
$k$, so that a large--scale ($k\to 0$) perturbation in the monolayer
density would relax $\propto \exp(-k^2 D_0 \beta_0 t)$, exhibiting a
Gaussian tail in the distribution of particles in real space. The 3D
hydrodynamic interactions lead to a value $\beta_{-1} \neq 0$ in
\eq{def_Dk}, signaling anomalous diffusion beyond a characteristic
length scale $\lhyd = \beta_0/\beta_{-1}$.  This divergence of $D(k\to
0)$ leads, in the real--space particle distribution, to an algebraic
decay $\propto x^{-3}$ with the distance $x$ from a density
perturbation.

These conclusions rely on constraining the particles to a
  monolayer, identified conventionally with the plane $z=0$. This is
the simplest model of an actual experimental configuration, where the
particles are trapped at a fluid interface by wetting forces or are
forced to stay within a plane by the effect of a strong external
potential (e.g., a sheet--like trap created by optical tweezers, the
gravitational field if the particles are sufficiently heavy 
to reside in a bottom layer \cite{Zahn:1997}, or the electrostatic attraction
to an interface \cite{LvHS07}). The goal of this work is to relax the 
strong--confinement assumption by considering the quasi--monolayer
  configuration. We will allow for a more realistic model in which
the position of the particles can fluctuate in the $z$--direction, so
that the associated dynamics is truly 3D, and will address how the
large--scale anomalous diffusion emerges from the underlying 3D normal
diffusion. If the quasi--monolayer is characterized by a
small thickness $\confell$, then the main result of our
analysis is that the diffusion is normal on scales below $\confell$,
regardless of the presence of hydrodynamic interactions, but the
anomalous--diffusion scenario is observed on scales above $\confell$.

The article is arranged as follows: in Sec.~\ref{sect:acd} the
theoretical model is described, including a brief review of the
general framework and the emergence of anomalous diffusion. The
special case of a harmonic trap in the $z$--direction is studied with
detail in the linear approximation in density perturbations.
Sec.~\ref{sect:results} presents results from truncated Stokesian Dynamics
  simulations of particles in the harmonic trap as well as from numerical
  solutions of 
  the corresponding dynamical evolution equation beyond the linear
  approximation.
The last Section
summarizes our conclusions. The Appendices collect the more technical
parts of the work.

\section{Theoretical model}
\label{sect:acd}

For the theoretical description of the dynamics of colloids, the
simplifying assumption can be done that the evolution occurs in the
overdamped regime and that the ambient flow can be described with the
time--independent Stokes equation, i.e., small Reynolds number and
instantaneous adjustment of the flow to the particle
configuration. This is usually a good approximation for the diffusive
dynamics because the time scale of change of the conserved field
``particle density'' diverges as the spatial extension of a density
perturbations is taken arbitrarily large (but see Ref.~\cite{Dom15} for a
discussion of how the anomalous diffusion is affected by allowing for
the dynamical evolution of the ambient flow). Under these
approximations, the hydrodynamic interactions mediated by the ambient
fluid can be taken completely into account by means of the mobility
matrix $\mathcal{M}_{ij}(\{\bx\})$, a $3\times 3$ matrix for each
particle pair $(i,j)$ \cite{Dhon96,KiKa91}.
It depends on the position $\{\bx\} = (\bx_1,\dots, \bx_N)$ of all the
$N$ particles forming the colloid and implicitly on their shape and
size through the boundary conditions that they impose on the ambient
flow. The physical meaning of this matrix is that the velocities
$\bv_i$ of the particles are determined by the forces $\mathbf{f}_j$
acting on them as
\begin{equation}
  \label{eq:mobility}
  \bv_i = \sum_{j=1}^N \mathcal{M}_{ij} \cdot \mathbf{f}_j .
\end{equation}
Correspondingly, the time evolution of the probability distribution
$P(\{\bx\}, t)$ of a configuration of particles at temperature $T$ is
described by the Smoluchowski equation~\cite{Dhon96,Allen:1987}:
\begin{subequations}
 \label{eq:smoluHydro}
\begin{equation}
 \frac{\partial P}{\partial t} = \sum_{ij} \nabla_{\vect x_i} \cdot
 \left( \mathcal{M}_{ij} \cdot \boldsymbol{\Phi}_j\right) ,
 \end{equation}
 \begin{equation}
   \boldsymbol{\Phi}_j(\{\bx\}) := 
   \kt \nabla_{\vect x_j} P(\{\bx\}) +  P(\{\bx\})\, \nabla_{\vect x_j} U(\{\bx\}) ,
 \end{equation}
\end{subequations}
with the potential energy 
\begin{equation} 
  \label{eq:potenergy}
  U(\{\bx\}) = U^\mathrm{int}(\{\bx\}) + \sum_{k=1}^N V(\vect{x}_k),
\end{equation}
consisting of an internal part $U^\mathrm{int}$ describing the
interparticle forces, and a contribution $V$ by an external
single--particle potential (in particular, the potential confining the
particles to the plane $z=0$). Equivalently, the dynamical evolution
for the individual particle trajectories $\bx_i(t)$ can be
described by the associated Langevin equation \cite{Allen:1987},
\begin{subequations}
  \label{eq:posLang}
\begin{equation}
  \label{eq:posLangHydro}
  \dot\bx_i = \sum_{j=1}^N \left[ - \mathcal{M}_{ij} \cdot
    \nabla_{\bx_j} U + \kt \nabla_{\bx_j}\cdot\mathcal{M}_{ij}
  \right]+ \bet_i ,
\end{equation}
in terms of a configuration--dependent Gaussian noise with zero mean
and variance
\begin{equation}
  \label{eq:posLangHydro_noise}
  \langle\eta_{i,\alpha}(\{\bx\}, t)\, \eta_{j,\beta}(\{\bx\}, t') \rangle = 2 \kt
  \left[\mathcal{M}_{ij}(\{\bx\})\right]_{\alpha\beta}\, \delta(t-t') 
\end{equation}  
\end{subequations}
(the Greek subindices refer to the components of the vectors and
  tensors.)  Particularly relevant is the driving force proportional
to the divergence of the mobility: it vanishes in bulk, i.e., when the
particles can be distributed in the volume, because the Stokes flow is
incompressible. However, in the partial confinement configuration, the
particles, but not the fluid are constrained to a plane, and this term
has the form of a nonvanishing 2D divergence of a 3D mobility matrix
(see, c.f., \eq{eq:langevinInplane}).

The collective (large scale, long time) dynamics of a colloid is
described by the evolution of the one-particle density distribution,
\begin{equation}
  \label{eq:density}
  \rho(\bx, t) = \int d^3\bx_2\dots d^3\bx_N \; P(\bx_1=\bx, \bx_2, \dots
  \bx_N, t) .
\end{equation}
One cannot derive a closed equation for $\rho(\bx,t)$ from the
Smoluchowski equation~(\ref{eq:smoluHydro}) without the introduction
of further approximations because of the multiparticle dependence of
both the mobility matrix $\mathcal{M}_{ij}(\{\bx\})$ and the potential
energy 
$U(\{\bx\})$. The simplest approximation, which
  will be adopted in this work, is to consider the dilute limit. For
the mobility matrix, this implies truncating its expansion at the
two--particle level and retaining the asymptotically dominant
contributions for large interparticle separations ($\mathcal{I}$ is
the identity matrix),
\begin{equation}
  \label{eq:nbodyM}
  \mathcal{M}_{ij}(\{\bx\}) = 
  \Gamma \delta_{ij} \mathcal{I} + \Gamma (1 - \delta_{ij})\,
  \bom (\vect x_i - \vect x_j) ,
\end{equation}
in terms of the single--particle mobility
\begin{equation}
  \label{eq:Gamma}
  \Gamma =\frac{1}{3\pi\eta\sigh} 
\end{equation}
(for our case of spherical particles of diameter $\sigh$ inmersed in a
fluid of viscosity $\eta$), and the Oseen tensor
\begin{equation}
  \label{eq:OS}
  \bom (\vect x) = \frac{3}{8} \frac{\sigh}{x} 
  \left( \mathcal{I}  + \frac{\vect x \vect x}{x^2} \right) .
\end{equation}
Effectively, one is taking into consideration only the longest ranged
contribution of the hydrodynamic interactions.

For the potential energy, 
the dilute limit approximation
  means $U^\mathrm{int}(\{\bx\}) = 0$ in \eq{eq:potenergy}, so that
  the particles do not interact directly with each other (``ideal
  gas'' approximation). With this approximation and \eq{eq:nbodyM},
  one can obtain from \eq{eq:smoluHydro} the following (nonlinear) 
evolution equation for the one--particle density:
\begin{subequations}
  \label{eq:cont}
\begin{equation}
  \label{eq:rho}
  \frac{\partial\rho}{\partial t} = D_0 \nabla_\bx^2 \rho -
  \nabla_\bx\cdot \left[ 
    \rho \left( \bu - \Gamma \nabla_\bx V \right)
  \right] ,
\end{equation}
where (see \eq{eq:Gamma})
\begin{equation}
  D_0 = \Gamma \kt = \frac{\kt}{3\pi\sigh\eta},   
\end{equation}
and
\begin{align}
  \label{eq:uambient}
  \bu(\bx) 
  & = 
   \int d^3\bx' \; \left[ D_0 \nabla_{\bx'} \rho (\bx') 
    - \Gamma \rho (\bx') \nabla_{\bx'} V(\bx') 
    \right] \cdot \bom (\bx-\bx') , 
    \nonumber \\
  & = 
   - \Gamma \int d^3\bx' \; \rho (\bx') [\nabla_{\bx'}
    V(\bx')] \cdot \bom(\bx-\bx') ,
\end{align}
\end{subequations}
after integrating by parts and accounting for the incompressibility
constraint $\nabla_\bx\cdot \bom (\bx)=0$. Since the Oseen tensor is
the Green function of the Stokes equation in an unbounded volume, one
can interpret the field $\bu(\bx)$ as the ambient flow induced by the
external forces acting on the particles. Then, \eq{eq:rho} describes
the evolution of the particle density due to Brownian motion and the
simultaneous drag by the ambient flow and the external force.

Although the dilute limit approximation is useful for the purpose of
this work, it can be relaxed. Thus, in order to account for the effect
of direct interparticle forces, local equilibrium approximations for
the potential energy landscape are customary. For instance, the
so-called dynamical density functional theory~\cite{MaTa99} and its
extension to include the effect of the hydrodynamic
interactions~\cite{Rex:2009,Donev:2014,Bleibel:JPCM15}.  The relevant
result is that, for the large--scale dynamics, the effect of the
short--ranged interparticle forces shows up as a (possibly
density--dependent) change in the numerical value $D_0$ of the
diffusion coefficient. Thus, it is not expected that the inclusion
  of direct interparticle forces will alter the qualitative picture,
  particularly that concerning anomalous diffusion in the monolayer
  configuration. This latter expectation is actually confirmed by
  numerical simulations of monolayers composed of interacting
  particles (capillary monopoles \cite{Bleibel:SMC14}, hard spheres
  \cite{GNK16}, Lennard--Jones particles \cite{PPD17}; see also the
  discussion in Ref.~\cite{Bleibel:JPCM15}).

Similarly, for monolayers formed at or close to a fluid interface, a
more realistic description of the mobility matrix is possible that
accounts for the different values of the fluid viscosity and the
particle positioning off the interface
\cite{Jones:1975,AdBl78,LoLa14}. Nevertheless, it turns out that
  the dominant far--field behavior is given again by the Oseen
  tensor, but with the viscosity $\eta$ in \eq{eq:nbodyM} replaced by
the arithmetic mean of the fluid viscosities\footnote{The Oseen
  tensor, decaying as $1/x$, is associated to a so-called
  ``Stokeslet'' \cite{KiKa91}.  The corrections thereof can be written
  as combinations of ``stresslets'' and ``rotlets'' (decaying as
  $1/x^2$) and higher--order terms~\cite{AdBl78,LoLa14}.}.
Therefore, no qualitative change in the large--scale behavior is expected
either.

\subsection{The partial confinement configuration}

\begin{figure}[ht]
  \begin{center}
    \epsfig{file=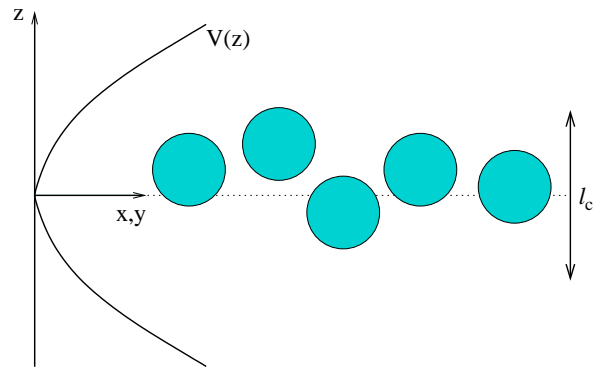,width=0.9\linewidth}
  \end{center}
  \caption{Schematic view of a setup of colloidal particles
      confined near the plane $z=0$ by an external potential that
      restricts the motion of the particles to a quasi--monolayer
      of thickness $\confell$}.    
  \label{fig:setup} 
\end{figure}

One can consider the particular case that the single--particle
external potential $V$ depends only on the $z$--coordinate and has the
proper form to force the confinement of the particles within a sheet
about $z=0$ of width $\confell$, see Fig.~\ref{fig:setup}. A
good example is a harmonic potential,
\begin{equation}
  \label{eq:harmonic_pot}
  V = \kt \left(\frac{z}{\confell}\right)^2 ,
\end{equation}
which will be developed in detail below. Before, however, we note some
remarks valid beyond this specific form of the potential. One can
address two limiting behaviors. In the limit $\confell\to\infty$
(absence of confinement), one effectively has $V \to 0$ and the
Smoluchowski equation~(\ref{eq:smoluHydro}) describes the dynamics
when the particles can explore the whole 3D volume. In the
approximated model described by \eqs{eq:cont}, one recovers normal
diffusion with the single--particle diffusion coefficient $D_0$. One
notices that the approximations leading to these equations are too
simple to capture any effect by the hydrodynamic interactions that
would induce a renormalization of the value of $D_0$. These appear
when short--distance effects are taking into account, like in the more
realistic hard--sphere model (see, e.g., Ref.~\cite{Mazur}).
Alternatively, these corrections could be incorporated, in the context
of a large--scale, long--time model for the dynamics of the
one-particle density, in the form of an effective density--dependence
of the single--particle mobility~\cite{Nozi87,Feld88,Noet89,Lhui90},
which must then be interpreted as a rheological parameter.

The situation is different, however, in the other limiting case,
$\confell\to 0$, which describes perfect confinement of the particles
to a monolayer in the plane $z=0$. We introduce the in--plane
  coordinate $\br=(x,y)$ such that
\begin{equation}
  \bx=\br + z\be_z, 
\end{equation}
and let $\bk$ denote the wavenumber vector for the 2D Fourier
transform with respect to the in--plane coordinate $\br$. The
probability distribution now has the structure
\begin{equation}
  \label{eq:confinedP}
  P(\{\bx\}) = P^\mathrm{(2D)} (\{\br\}) \prod_{i=1}^N \delta (z_i) ,
\end{equation}
in terms of the in--plane projected probability distribution
$P^\mathrm{(2D)} (\{\br\})$. Furthermore, there is no vertical particle
current, i.e., $\be_z\cdot\boldsymbol{\Phi}_i=0$ in the Smoluchowski
equation~(\ref{eq:smoluHydro}), and the particle distribution in the
$z$--direction is always in equilibrium regardless of the dynamical
state of the in--plane distribution. Therefore, upon integrating the
Smoluchowski equation over the $z$--coordinates of the particles, one
arrives at
\begin{equation}
  \label{eq:inplanesmoluHydro}
  \frac{\partial P^\mathrm{(2D)}}{\partial t} = \sum_{ij} \nabla_{\br_i} \cdot
  \left\{ \mathcal{M}_{ij} \cdot \left[ \kt \nabla_{\br_j} P^\mathrm{(2D)}
      +  P^\mathrm{(2D)} \nabla_{\br_j} U^\mathrm{int} \right] \right\} ,
\end{equation}
that is, the Smoluchowski equation for the 2D dynamics in the plane
$z=0$, where any  
reference to the confining potential has disappeared.  
However, the mobility matrix $\mathcal{M}_{ij}(\{\br\})$ still describes a 3D
flow 
(although evaluated at the plane $z=0$), and this dimensional mismatch
leads to the anomalous diffusion. The associated Langevin equation
  for the 2D trajectories $\br_i(t)$ of the particles in the monolayer
  has the form 
\begin{subequations}
\begin{equation}
  \label{eq:langevinInplane}
  \dot\br_i = 
  \frac{1}{8\pi\eta} \sum_{j=1, j\neq i}^N
  \frac{\br_i - \br_j}{|\br_i - \br_j|^3} + \boldsymbol{\xi}_i ,
\end{equation}
\begin{multline}
  \langle \xi_{i,\alpha}(\{\br\}, t) \xi_{j,\beta}(\{\br\}, t')
  \rangle = 2 D_0 \left[ 
    \delta_{ij} \delta_{\alpha\beta} 
    \right. \\
    \left. + ( 1 - \delta_{ij} ) \omega_{\alpha\beta} (\br_i-\br_j)
    \right] \delta(t-t') .
\end{multline}
\end{subequations}
when \eq{eq:posLangHydro} is projected onto the monolayer plane
  with the approximations~(\ref{eq:nbodyM}) and $U^\mathrm{int}=0$.
The force term in \eq{eq:langevinInplane} follows from the
  observation that
  $\nabla_{\br_j}\cdot\mathcal{M}_{ij}\neq 0$ since the in-plane
  component of the 3D ambient flow \emph{is not} 2D incompressible in
  general, and leads to a force term proportional to
  $\nabla_\br\cdot\omega(\br)$ which is formally identical to a
  Coulombic repulsion, 
  the ultimate cause of the superdiffusive behavior.

The corresponding equation for the 2D particle density field
$\rho^\mathrm{(2D)}(\br)$ is
\begin{subequations}
  \label{eq:inplanecont}
\begin{equation}
  \label{eq:rho2D}
  \frac{\partial\rho^\mathrm{(2D)}}{\partial t} = D_0 \nabla_\br^2
  \rho^\mathrm{(2D)} - 
  \nabla_\br\cdot ( \rho^\mathrm{(2D)} \bu ) ,
\end{equation}
\begin{eqnarray}
  \bu(\br) & = & D_0 \int d^2 \br' \; \left[ \nabla_{\br'} \rho^\mathrm{(2D)}
    (\br') \right] 
  \cdot \bom(\br-\br') \nonumber \\
  & = & D_0 \int d^2 \br' \; \rho^\mathrm{(2D)} (\br') 
    \; \nabla_{\br} \cdot \bom(\br-\br').
\end{eqnarray}
\end{subequations}
Again, any overt signature of the confining potential has disappeared,
and \eq{eq:rho2D} describes the dynamical evolution driven by the
  in--plane Brownian diffusion and the drag by the in--plane component
  $\bu(\br)$ of the ambient flow. However, the field $\bu(\br)$ is
induced by the in--plane particle current, which is now a relevant
source because, unlike in the derivation of \eq{eq:uambient},
$\nabla_\br\cdot\bom(\br) \neq 0$.
As shown in Refs.~\cite{Bleibel:SMC14,Bleibel:JPCM15}, the
linearization of this equation for small perturbations about a
homogeneous in--plane density $\rho^\mathrm{(2D)}_0$ yields a wavenumber
dependent diffusion coefficient (see \eq{def_Dk}),
\begin{equation}
  \label{eq:anomD}
  \frac{D(k)}{D_0} =  1 + \rho^\mathrm{(2D)}_0 \frac{\bk}{k}\cdot
  \mathrm{FT}[\bom] \cdot \frac{\bk}{k}
  = 1 + \frac{1}{\lhyd k} ,
\end{equation}
where $\mathrm{FT}[\bom]$ denotes the 2D Fourier
transform of the 3D Oseen tensor and
\begin{equation}
  \label{eq:Lhydro_gen}
  \lhyd := \frac{4}{3 \pi \sigh \rho^\mathrm{(2D)}_0} 
\end{equation}
is a characteristic length scale.

\subsection{Harmonic confinement: linear theory}
\label{subsect:HCLT}

The goal is to investigate the transition from one limiting case to
the other, with emphasis on  
the quasi--monolayer configuration. For this purpose we address in detail the
linearized theory for the equation~(\ref{eq:cont}) with a harmonic confining
potential, see \eq{eq:harmonic_pot}. The equilibrium state is given by
the Boltzmann distribution,
\begin{equation}
  \label{eq:Boltzmann}
  \rho_\mathrm{eq}(z) = \frac{\rho_0^\mathrm{(2D)}}{\sqrt{\pi}\confell} \; 
  \mathrm{e}^{-V(z)/\kt} ,
\end{equation}
where $\rho_0^\mathrm{(2D)}$ is the projected 2D number density,
\begin{equation}
  \rho_0^\mathrm{(2D)} = \int_{-\infty}^{+\infty} dz\; \rho_\mathrm{eq}(z) .
\end{equation}
 Any particle distribution can be expressed as
\begin{equation}
  \label{eq:rhopert}
  \rho(\vect{r},z,t) = 
  \rho_\mathrm{eq}(z) \left[ 1+\vare(\vect{r},z,t) \right] .
\end{equation}
When the model equation~(\ref{eq:cont}) is linearized with respect to
the small perturbation $\vare$ one obtains an integro--differential
equation for the evolution of the fluctuations (see
App.~\ref{app:theory}):
\begin{subequations}
  \label{eq:lindiff}
\begin{equation}
  \label{eq:lineps}
  \frac{\partial \vare}{\partial t} = 
  D_0 \left[ \nabla^2 + \frac{\partial^2}{\partial z^2} \right] \vare
  + \frac{\Gamma}{D_0} \frac{d V}{d z} 
  \left[ - D_0 \frac{\partial \vare}{\partial z}
    + \be_z\cdot\bu 
  \right], 
\end{equation}
\begin{multline}
  \label{eq:linu}
  \bu(\br,z,t) = - \Gamma \int d^2\br^{\prime} \int_{-\infty}^{+\infty} dz' \; 
  \rho_\mathrm{eq}(z^{\prime}) \frac{d V}{d z^{\prime}}(z') \vare(\br^{\prime},z^{\prime},t)
  \, \\ \times \be_z \cdot \bom(\br-\br^{\prime} + (z-z^{\prime})\be_z ).
\end{multline}
\end{subequations}
By introducing the Fourier transform for the in--plane
$\br$--dependence and an expansion in Hermite polynomials $H_n$ for
the vertical $z$--dependence, one can write
\begin{equation}
  \label{eq:Hermite}
  \vare(\br,z,t) = \int \frac{d^2 \bk}{(2\pi)^2} \;
  \mathrm{e}^{i\bk\cdot\br}  
  \; \sum_{n=0}^{\infty} H_n \left( \frac{z}{\confell} \right) \, c_n(\bk,t) .
\end{equation}
Particularly relevant is the coefficient $c_0(\bk,t)$, that describes
the Fourier transform of the vertically integrated density profile,
which is the 2D density distribution in the partial confinement limit:
\begin{subequations}
  \label{eq:inplanerho}
\begin{equation}
  \int_{-\infty}^{+\infty} dz\; \rho(\br,z,t) = \rho_0^\mathrm{(2D)}
  \left[ 1 +   \delta_0(\br, t)  \right] ,
\end{equation}
with
\begin{equation}
  \delta_0(\br, t) := 
  \int \frac{d^2 \bk}{(2\pi)^2} \; \mathrm{e}^{i\bk\cdot\br} c_0 (\bk,t) .
\end{equation}
\end{subequations}
The linearized equation~(\ref{eq:lindiff}) then becomes a set of
linear equations for the coefficients $c_n(\bk,t)$ (see
App.~\ref{app:theory}):
\begin{subequations}
  \label{eq:fullc}
\begin{equation}
  \label{eq:c}
  \frac{\confell^2}{D_0} \frac{d c_n(\bk,t)}{d t} = - [ (\confell k)^2 + 2 n]
  c_n(\bk,t)  
  + \frac{\psi_n(\bk,t)}{\lhyd k} ,
\end{equation}
\begin{equation}
  \label{eq:psi}
  \psi_n(\bk,t) = - \frac{1}{\pi \, n! \, 2^n} 
  \sum_{m=0}^{\infty} \Omega_{nm}(\confell k) c_m(\bk,t) .
\end{equation}
\end{subequations}
Here, $\lhyd$ is given by \eq{eq:Lhydro_gen} and the term
$\psi_n$ encodes the effect of the (long--ranged part of the)
hydrodynamic interactions. 
The dimensionless coefficients $\Omega_{nm}$ are computed in
App.~\ref{app:psi}. Of particular relevance is that they are symmetric
under the exchange of the indices $n,m$ and vanish when they have
different parity. As a consequence, the equations~(\ref{eq:fullc})
actually form two uncoupled sets of equations: the set for $c_n$, $n$
odd, describes particle distributions that are asymmetric in $z$ and
whose evolution is driven both by diffusion and the net force exerted
by the confining potential. Therefore, we limit ourselves for
simplicity to symmetric perturbations in the following, i.e.,
$\vare(\br,-z,t) = \vare(\br,+z,t)$, so that the net external force
vanishes and we have to 
consider only the dynamics of the coefficients
$c_n$ with $n$ even. 

The relatively simple structure of \eqs{eq:fullc} describes the diffusive
relaxation of the modes on a time scale controlled by the length
$\confell$, and the coupling mediated by the hydrodynamic interactions
with a strength controlled by the length scale $\lhyd$.
Two regimes are particularly interesting:

\noindent
(i) For the very small in--plane scales, $\confell k \gg 1$ and
$\lhyd k \gg 1$, neither the confining potential nor the
hydrodynamic interactions affect the dynamical evolution appreciably:
3D normal diffusion is recovered because the modes evolve on a time
scale $\sim 1/D_0 k^2$ while the term $\psi_n$ is strongly suppressed
(in addition to the $1/(\lhyd k)$ prefactor, it is
$\Omega_{nm}(\confell k\to \infty) \sim 1/(\confell k)$, see
App.~\ref{app:psi}).

\noindent
(ii) For in--plane scales much larger than the thickness of the
quasi--monolayer, $\confell k \ll 1$, one can recover the scenario
originally studied in Ref.~\cite{Bleibel:SMC14}, as well as derive the
leading correction for a nonvanishing thickness $\confell$.
On the one hand, the mode $c_0$, associated to the
conserved 2D density distribution, is a slow variable, with a
characteristic time scale vanishing when $\confell k \to 0$. On the
other hand, the modes $c_{n\geq 2}$ relax on the fast time scale $\sim
2 n D_0/ \confell^2$, signaling the onset of the Boltzmann
distribution in the vertical direction. Therefore, the effective
dynamics of $c_0$ on the slow time scale can be computed approximately
by an adiabatic elimination of the fast modes from its equation: the
modes $c_{n\geq 2}$ decay to their stationary value at fixed $c_0$ and
get ``enslaved'' to the dynamical evolution of the latter. This
procedure is detailed in App.~\ref{app:adiabatic}; the final result is
\begin{subequations}
  \label{eq:mainresult}
  \begin{equation}
    \label{eq:dync0}
    \frac{d c_0}{d t} = - k^2 D(k) c_0 , 
  \end{equation}
  \begin{equation}
    \label{eq:longD}
    \frac{D(k)}{D_0} - 1 \approx 
    \frac{1}{\lhyd k}  
    - \sqrt{\frac{8}{\pi}}\; \frac{\confell}{\lhyd}
    - \frac{1}{2}\left(\frac{\confell}{\lhyd}\right)^2 ,
  \end{equation}
\end{subequations}
where $D(k)$ is derived from an expansion in the small parameter
$\confell k$. Therefore, the hydrodynamic interactions give rise to
anomalous diffusion for the large in---plane scales satisfying $\lhyd
k \ll 1$, in agreement with \eq{eq:anomD} for the case $\confell=0$.
In addition, there is a finite renormalization of the diffusion
coefficient $D_0$ for finite values of $\confell$, but this effect will
be hardly observable: it is quantitatively relevant only when $\lhyd$
is of the order of $\confell$, in which case it will be $\lhyd k \sim
\confell k \ll 1$, and the anomalous--diffusion effect dominates anyway.

In conclusion, the crossover from 3D normal diffusion to 2D anomalous
diffusion occurs smoothly as one shifts the attention from the
smallest to the largest scales, with the two length scales $\confell$
(width of the confining potential) and $\lhyd$ (onset of anomalous
diffusion) controlling this transition. Our detailed analysis above
for the case $\confell\ll \lhyd$ reveals the following
  hierarchy of dynamical regimes in wavenumber:
\begin{center}
\begin{tabular}[t]{cc}
  (I) & bulk (3D) normal diffusion \\
  & if $\lhyd^{-1}\ll \confell^{-1} \ll k$ \\
  & \\
  (II) & in--plane (2D) normal diffusion \\
  & if $\lhyd^{-1} \ll k \ll \confell^{-1}$ \\
  & \\
  (III) & in--plane (2D) anomalous diffusion \\
  & if $k \ll \lhyd^{-1}\ll \confell^{-1}$
\end{tabular}
\end{center}
Alternatively, one gets the following scenario in terms of the length
scale $r$ of observation: at the smallest scales ($r \ll \confell \ll
\lhyd$) the particle distribution diffuses normally inside the
quasi--monolayer (regime I); when observed at the intermediate
scales ($\confell \ll r \ll \lhyd$), the particle distribution already
appears as a perfect monolayer and diffuses normally in the monolayer
plane (regime II); and at the largest scales ($\confell \ll
\lhyd \ll r$), the monolayer diffusion is anomalously fast
(regime III).

\section{Numerical calculations}
\label{sect:results}

\subsection{Setup and numerical methods}
To further illustrate and complement the results of linear theory
for the harmonic confining potential, we performed truncated
Stokesian Dynamics (tSD) simulations on the one hand, and solved
numerically the density evolution equation (DEE) (\ref{eq:cont})
on the other hand. In both cases the particles were modeled as
  spheres of diameter $\sigma_H=20\;\mu\mathrm{m}$ and the fluid was
  taken at room temperature ($T=25^o\mathrm{C}$) with the viscosity of
  water $\eta=10^{-3}\;\mathrm{N\,s/m}^2$. 

The initial particle distribution was constructed as the
  superposition
\begin{equation}
  \label{eq:initrho}
  \rho(\bx) = \rho_\mathrm{eq}(z) + \Delta\rho(r, z) ,
\end{equation}
where the background density $\rho_\mathrm{eq}(z)$ is given by
  \eq{eq:Boltzmann}, which is modified by the radially
  symmetric overdensity $\Delta\rho(r, z)$. For the latter,
we investigated two cases:
\begin{itemize}
\item[(i)] A planar overdensity which is equilibrated in the
  $z$--direction but is constant and nonzero on a disk of radius $R$
  in the $x$--$y$--plane,
  \bea 
  \label{eq:initialdisk}
  \Delta\rho(r, z) = \rho_\mathrm{eq}(z) \; A_0 \,\Theta(R-r) \;. 
  \eea
  With this setup we will exemplify the behavior in regimes II and
  III defined above.
\item[(ii)] A narrow and isotropic peak of width $\lG = \confell/10$
  which is centered at $(r,z)=(0,0)$,
  \bea
  \label{eq:initialpeak}
   \Delta\rho(r, z) =  \rho_{\rm G} \exp\left( -
    \frac{r^2+z^2}{\lG^2}\right) \;, 
  \eea
  with the choice $\rho_{\rm
      G}=(A_0/\sqrt{\pi})(\rho_0^\mathrm{(2D)} R^2/\lG^3)$, so that
    the number of particles in the overdensity is the same as in the
    planar one~(\ref{eq:initialdisk}). This case will allow us to
    address the behavior in the regime I defined above.
\end{itemize} 
For case (i) we have obtained results from both tSD simulations and
the solution of the DEE, while case (ii) has been investigated
with the DEE only. 

The tSD simulations solve the evolution of a collection of $N$
  particles whose dynamics is given by the Langevin
  equations~(\ref{eq:posLang}). No direct interaction is
  considered, $U^\mathrm{int}=0$, and the mobility matrix is
  approximated as in \eq{eq:nbodyM} with the pairwise hydrodynamic
  interaction given
  by the Rotne--Prager--Yamakawa tensor~\cite{RPY:6970}, 
  \begin{align}
  \label{eq:rp}
  \bom^{\rm RPY}(\vect x) & = \left\{ 
    \begin{array}{cc} 
      \displaystyle 
      \bom(\bx)
      + \frac{\sigh^3}{16 x^3} \left( 
        \one - \frac{3\vect x \vect x}{x^2}
        \right) , 
        & 
        (x\,>\,\sigh) ,
        \\
        & \\ 
        \displaystyle \left(1-\frac{9 x}{16 \sigh}\right) \one +
        \frac{3 x}{16 \sigh}\, \frac{\vect x \vect x}{x^2} ,
        &
        (x\, <\, \sigh).
      \end{array}
      \right.
  \end{align} 
  This tensor is regular at $x=0$ and positive definite, and therefore
  better suited for particle--based simulations than the Oseen tensor
  $\bom(\bx)$ (see \eq{eq:OS}). Beyond the dilute limit,
  a system of particles which do not interact directly (i.e.,
  ideal--gas behavior) but do it hydrodynamically (i.e., with a
  nonvanishing hydrodynamic radius $\sigh$ in
  \eq{eq:rp}) 
  can be realized physically by means of ``hairy'' particles
  consisting of a small solid core and a broad polymeric shell around
  it \cite{Lin:1995,Bleibel:JPCM15}.

The tSD simulations consisted of
  $N=1036$ spherical particles. 
  They were performed in a simulation box of extension $L=2000$
$\mu\rm{m}$ in the $x$--$y$--plane with periodic boundary conditions,
while unbounded in the $z$--direction since the confinement by the
harmonic potential effectively imposes a vanishing particle current at
infinity. The initial overdensity in \eq{eq:initialdisk} was simulated
with $N_{\rm disk}=188$ particles distributed uniformly within a
circular patch 
of radius $R=100$ $\mu$m and according to the equilibrium profile $\rho_\mathrm{eq}(z)$ in the vertical direction. The remaining $N_{\rm b}=848$
  particles were distributed similarly but over the whole planar extension of
  the system. 
Thus, the effective 2D background density is $\rho_0^\mathrm{(2D)}
  = N_\mathrm{b}/L^2 = 2.12\times 10^{-4}$ $\mu\mathrm{m}^{-2}$,
  corresponding to a packing fraction well in the dilute limit,
  $(\pi/4)\sigma_H^2 \rho_0^\mathrm{(2D)} \approx 0.07$. This 
  gives $A_0=N_\mathrm{disk}/(\pi R^2 \rho_0^\mathrm{(2D)}) = 28.2$ in
  \eq{eq:initialdisk}, which represents a large perturbation presumably
  beyond the scope of the linearized theory. The characteristic length scale
of anomalous diffusion~(\ref{eq:Lhydro_gen}) associated to this
initial configuration was $\lhyd=100$ $\mu$m.

The density evolution equation (DEE) (\ref{eq:cont}) is a nonlinear
integro--differential equation owing to the hydrodynamic term. We have
solved it using an Euler forward scheme for the time evolution. The
right hand side of \eq{eq:rho} was evaluated with Fourier
transforms. In the $x$--$y$--plane, radial symmetry was assumed and
the corresponding Fourier transforms could be evaluated on an
equidistant grid for $\ln r$ using Fast Hankel Transforms. In
$z$--direction, we used a Fast Fourier Transform on an equidistant
grid.  The box size in $z$--direction was chosen $L_z = 40.96 \;R$.
Since the system is periodic in $z$--direction through the use of the
Fast Fourier Transform, the results for density profiles in radial
direction in the $x$--$y$--plane are affected by periodic images for
$r \gtrsim L_z$.

For later use, we have expressed the thickness of the
  quasi--monolayer $\confell$
in terms of a dimensionless confinement parameter
\begin{equation}
  \label{eq:alpha}
  \alpha = \frac{\sqrt{2} \confell}{\sigh} .
\end{equation}
 
\subsection{Results: diffusion of a planar overdensity}
\label{sec:planar}

\begin{figure}[h]
  \begin{center}
    \epsfig{file=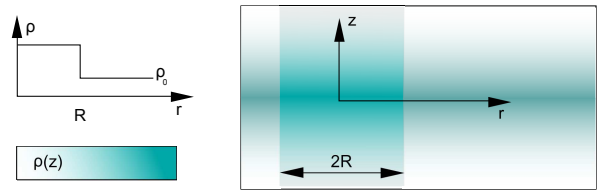,width=0.9\linewidth}
  \end{center}
  \caption{Schematic side view of the setup of the initial planar
    overdensity over the equilibrated background. The overdensity
    corresponds to the background density uniformly ``compressed'' to
    a disk of radius $R$.}
  \label{fig:setup_disk} 
\end{figure}

The setup for the planar overdensity given by
  Eqs.~(\ref{eq:initrho}, \ref{eq:initialdisk}) is shown in
Fig.~\ref{fig:setup_disk}.
This case is a straightforward extension of the planar overdensity
investigated earlier in strict 2D confinement \cite{Bleibel:SMC14} and
focuses on the effect of the finite width of the confining potential
in $z$--direction upon the dynamics in the
$x$--$y$--plane. 

We investigated the range of values $1.25 \leq \alpha \leq 10$ for
  the confinement parameter, corresponding to widths $\confell$
between 20 $\mu$m and 140 $\mu$m, i.e.~for the smallest width the
system is close to a monolayer and for the largest width the
$z$--extension of the initial overdensity is about as large as the extension
in the plane.  Of basic interest is the time evolution of the
$z$--averaged relative overdensity $\delta_0(r,t)$,
which corresponds to the inverse Fourier transform of the mode $c_0(k,t)$
(see \eqs{eq:inplanerho}). The scales in our setup satisfy
$\confell \lesssim R,\, L_{\rm hydro}$, so it can be conjectured
that the expansion of
 Sec.~\ref{subsect:HCLT} in Hermite modes is  
particularly well suited and fast converging for the density evolution
on scales $r \gg R, L_{\rm hydro}$.

\begin{figure}[ht!]
  \begin{center}
    \epsfig{file=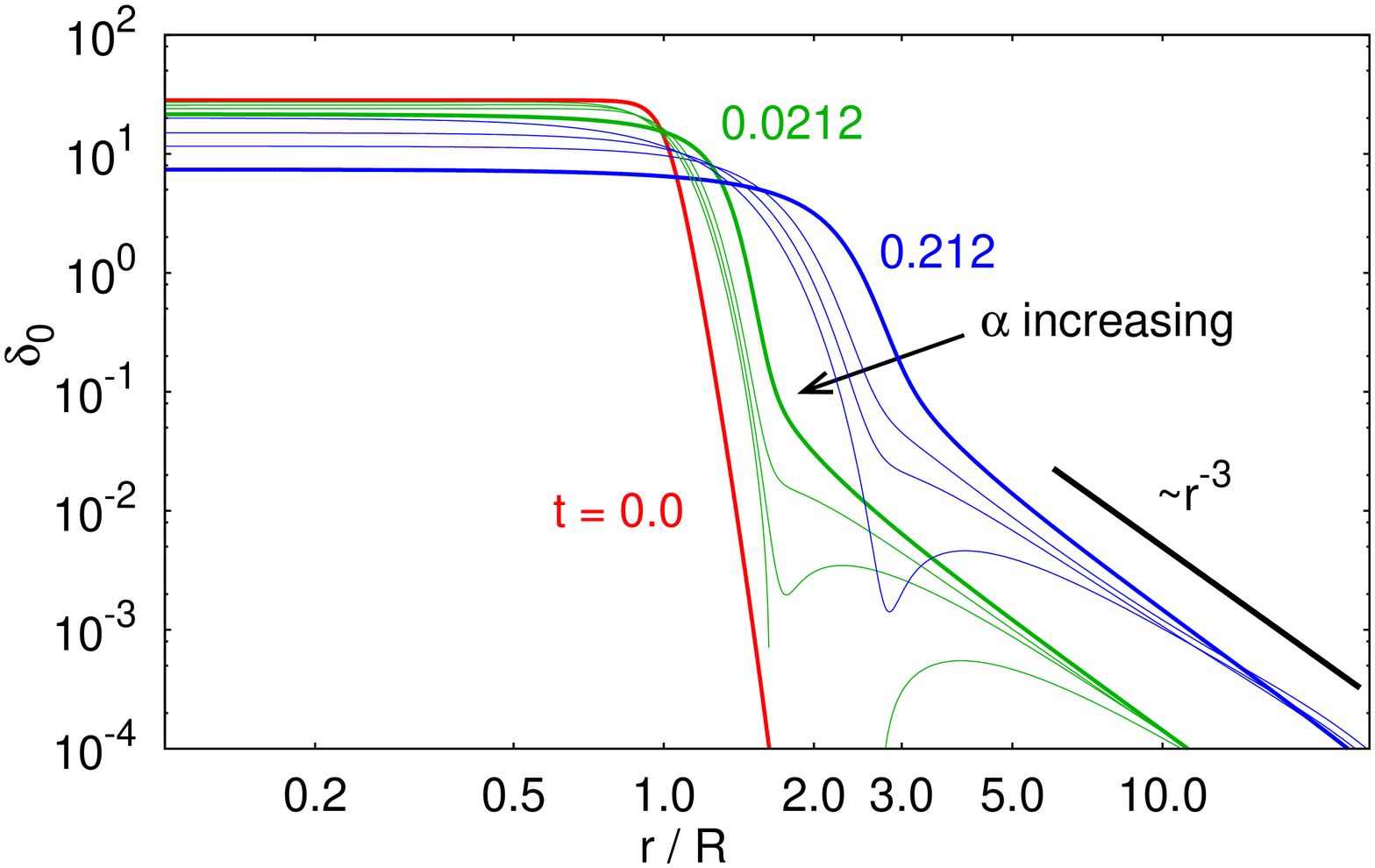,width=0.9\linewidth}
  \end{center}
  \caption{The $z$--averaged relative overdensity $\delta_0(r,t)$
    from the DEE solution, evaluated as a function of $r$ at two
    different times for different values of the confinement parameter: $\alpha=2.5$ (dashed), 5
      (dotted) and 10 (dashed--dotted). 
      Thick lines
    show the case of strict 2D confinement (\eqs{eq:inplanecont}) as a
    reference.  
    The initial
    overdensity is smoothened at the edge of the disk to avoid
    numerical artefacts.  The time unit is given by $1/(D_0
    \rho_0^\mathrm{(2D)})$.
  }
  \label{fig:plot1}
\end{figure}

In Fig.~\ref{fig:plot1}, DEE solutions for
$\delta_0(r,t)$ 
are shown for 
parameters $\alpha=2.5$, 5 and 10 at two different times $t/t_0 =
0.01$ and 0.1 together with the starting configuration.  (The time
unit $t_0=1/(D_0 \rho_0^\mathrm{(2D)})$ corresponds to the
characteristic Brownian diffusion time at which particles in the plane
reach their next neighbor in the background configuration.) As a
reference, the relative overdensity in the case of strict 2D
confinement (for the same two times, respectively) is
shown: 
The profile decays monotonously in space and shows the instantaneous onset of
the $1/r^3$--tail 
characterizing anomalous diffusion. The tail grows in magnitude with
time. For a finite thickness of the quasi--monolayer, the spatial density
profiles show the same 
asymptotic, anomalous decay which, however, sets in only at radial
distances larger than a critical one. This critical distance also
grows with the confinement parameter $\alpha$.  This finding has
a very straightforward interpretation: Only at distances beyond this
critical distance the $z$--confined overdensity appears to be
effectively 2D \emph{and} anomalously decaying. This is in full
accordance with the behavior in regime III characterized by the
singularity $\propto 1/k$ in the diffusion coefficient, see
  \eq{eq:anomD}, derived in the linearized theory.

\begin{figure}[ht!]
  \begin{center}
    \epsfig{file=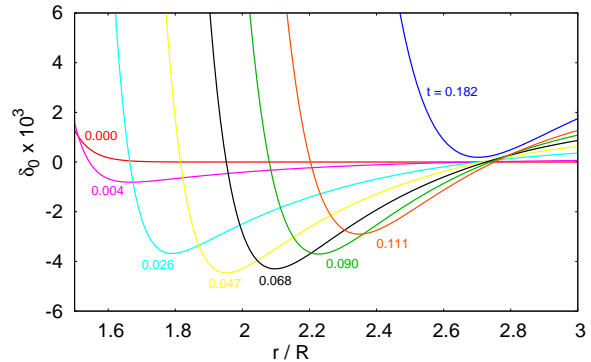,width=0.9\linewidth}
  \end{center}
  \caption{The $z$--averaged relative overdensity in the dip region for
    increasing times and 
    $\alpha=10.0$ obtained from the DEE solution. A depletion zone
    with 
    negative overdensity first develops between $1.6 < r/R < 2.8$ and 
    disappears
    for later times. }
  \label{fig:large_alpha_inset}
\end{figure}

For radial distances smaller than the critical one we enter regime II.
The diffusion of the disklike overdensity at small $r/R < 2$
becomes slower with increasing width of the confining potential. At
intermediate $r/R \approx 3$ a dip in the overdensity is formed before
the profile approaches the anomalous tail for large $r$. This dip
  is a consequence of the finite thickness of the quasi--monolayer
  because it is absent in simulations with strict 2D confinement,
  regardless of the initial extension of the planar overdensity. For
the largest width investigated ($\alpha=10$), the overdensity becomes
actually negative (i.e., there is a relative depletion) in the dip region.
We illustrate this with a time series of overdensity profiles for
$\alpha=10$ in Fig.~\ref{fig:large_alpha_inset} which magnifies the
dip region; the overdensity is negative for $1.6 < r/R <
2.8$. 
We observe that the diffusive motion of the edge of the disk
becomes slower with increasing $\alpha$; however, it always
triggers a hydrodynamic outbound flow at large distances,
responsible for the anomalous diffusion and the $1/r^3$--tail
(regime III). This mechanism drags particles away from the disk edge
at a faster rate than they can be replenished by normal diffusion
from the disk (regime II), thus developing an initial depletion zone
right at the outer edge of the disk, which becomes more conspicuous
for larger values of $\alpha$.
  Interestingly, this means that the 2D effective
  Green function for  
  the diffusive spread of the overdensity is no longer greater than $0$
  everywhere (whereas in strict 2D confinement it is). 

The initial planar overdensity is equilibrated in $z$--direction and thus only
the zeroth 
Hermite mode $c_0(k,t)$ is present. The dynamics, however, leads to a
distortion of 
the Gaussian $z$--dependence and to the appearance of higher Hermite modes
$c_2(k,t), c_4(k,t),...$. 
These initially grow in time, their strength reaches a maximum at a time $t
\sim \confell^2/D_0$ and 
then decays in time. Overall, these higher modes are always much smaller than
the leading, zeroth mode.  

\begin{figure}[ht!]
  \begin{center}
    \epsfig{file=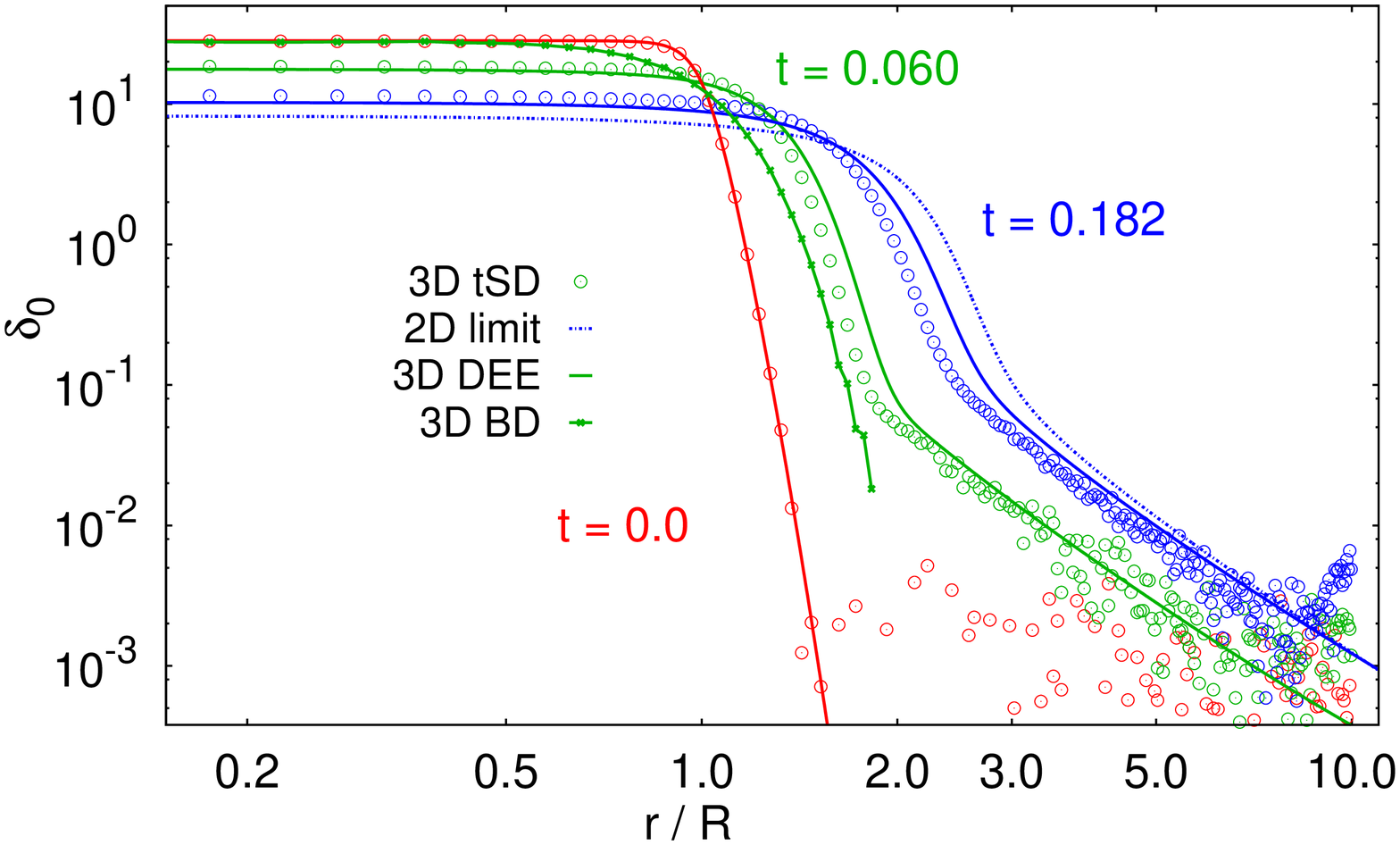,width=0.9\linewidth}
  \end{center}
  \caption{Density profiles (integrated in $z$--direction) obtained
    from 3D tSD simulations and 3D DEE solutions for the
      confinement parameter $\alpha=1.25$. The tSD data points
      were obtained from averaging over $150000$ simulation runs. For
    comparison we show the profiles of Brownian Dynamics simulations
    (3D BD, $\bu=0$ in \eq{eq:rho}) at time $t=0.06$ and the
      profiles from the numerical solution of \eqs{eq:inplanecont} (2D
      limit) at time
    $t=0.182$.
  }
  \label{fig:small_alpha}
\end{figure}

Next we compare DEE solutions to results from tSD simulations. For a small
value $\alpha=1.25$ we show 
in Fig.~\ref{fig:small_alpha}
DEE solutions and tSD results for the planar overdensity at two times and
compare them 
to the limit 
of strict 2D confinement  and to the limit of 3D Brownian diffusion. 
This value of $\alpha$ corresponds to a monolayer of thickness $\approx
\sigh=20$ 
$\mu$m. We 
observe that 
the temporal decay of the overdensity profile is qualitatively as in the
strict 2D case but the 
built-up 
of the anomalous tail is slower for intermediate $r/R$.
At large distances ($r/R>5$), the results from tSD and
the DEE solutions (both 2D and 3D) coincide. In both cases, the diffusion of
the edge  
of the disklike overdensity proceeds more slowly as compared to the strict 2D
case, but still faster than Brownian Dynamics. 

\begin{figure}[ht!]
  \begin{center}
    \epsfig{file=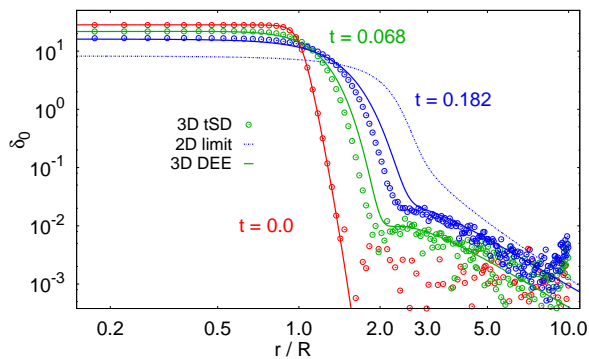,width=0.9\linewidth}
  \end{center}
  \caption{The same as Fig.\ref{fig:small_alpha} but for the
      confinement parameter $\alpha=5.0$.  }
  \label{fig:medium_alpha}
\end{figure}

We increase the width to $\alpha=5$ (the width of the confining potential is
about $3.5 
\sigh= 70$ $\mu$m). The tSD profiles clearly confirm the dip which we
discussed above 
for the DEE solutions, see Fig.~\ref{fig:medium_alpha}.
DEE
and tSD profiles agree at large distances, whereas for intermediate $r/R$ 
the evolution of the profile appears to proceed
more slowly in simulations, a fact that may be attributed to the rather small
box used in simulations.
However, the main features of the evolution are captured by both methods alike. 
These main features are 
the deviations from  the case of strict 2D confinement in the depletion zone
as well  
as the onset of anomalous diffusion at larger distances (regime III). 

\subsection{Results: diffusion of a narrow peak}
\label{sec:peak}

\begin{figure}[ht]
  \begin{center}
    \epsfig{file=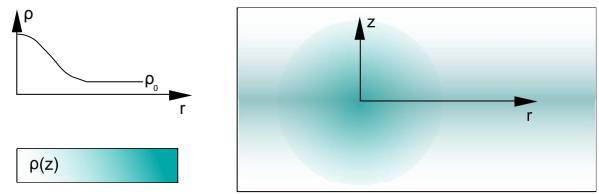,width=0.9\linewidth}
  \end{center}
  \caption{Schematic side view of the setup of an initial narrow,
    isotropic peak over the equilibrated background.  The number of
    particles in the peak is chosen to be the same as the one in the
    planar overdensity of Sec.~\ref{sec:planar}. }
  \label{fig:setup_blob} 
\end{figure}

The setup for the peaklike overdensity given by
  Eqs.~(\ref{eq:initrho}, \ref{eq:initialpeak}) is shown in
Fig.~\ref{fig:setup_blob}. 
The width of the confining potential was set to $\confell \approx 140$ $\mu$m
($\alpha=10$), so that 
  $\lG = \confell/10=14\;\mu\mathrm{m}$.
  For $\lG \to 0$, the time evolution corresponds to 
  the decay of a $\delta$--peak in the nonlinear DEE (\ref{eq:cont})
  (corresponding to a Green function for a linear DEE).
  This case allows us to study the transition from
  presumably normal diffusion at small lateral distances (regime
  I) to anomalous diffusion at larger distances and longer times
  (regime III).  In regime I, we are especially interested
  in the effect of hydrodynamics in the confined system on smaller
  length scales; therefore, we have compared the case with
    hydrodynamic interactions to the 3D, purely Brownian case
  ($\bu=0$ in \eq{eq:rho}).
 
\begin{figure}[ht!]
  \begin{center}
    \epsfig{file=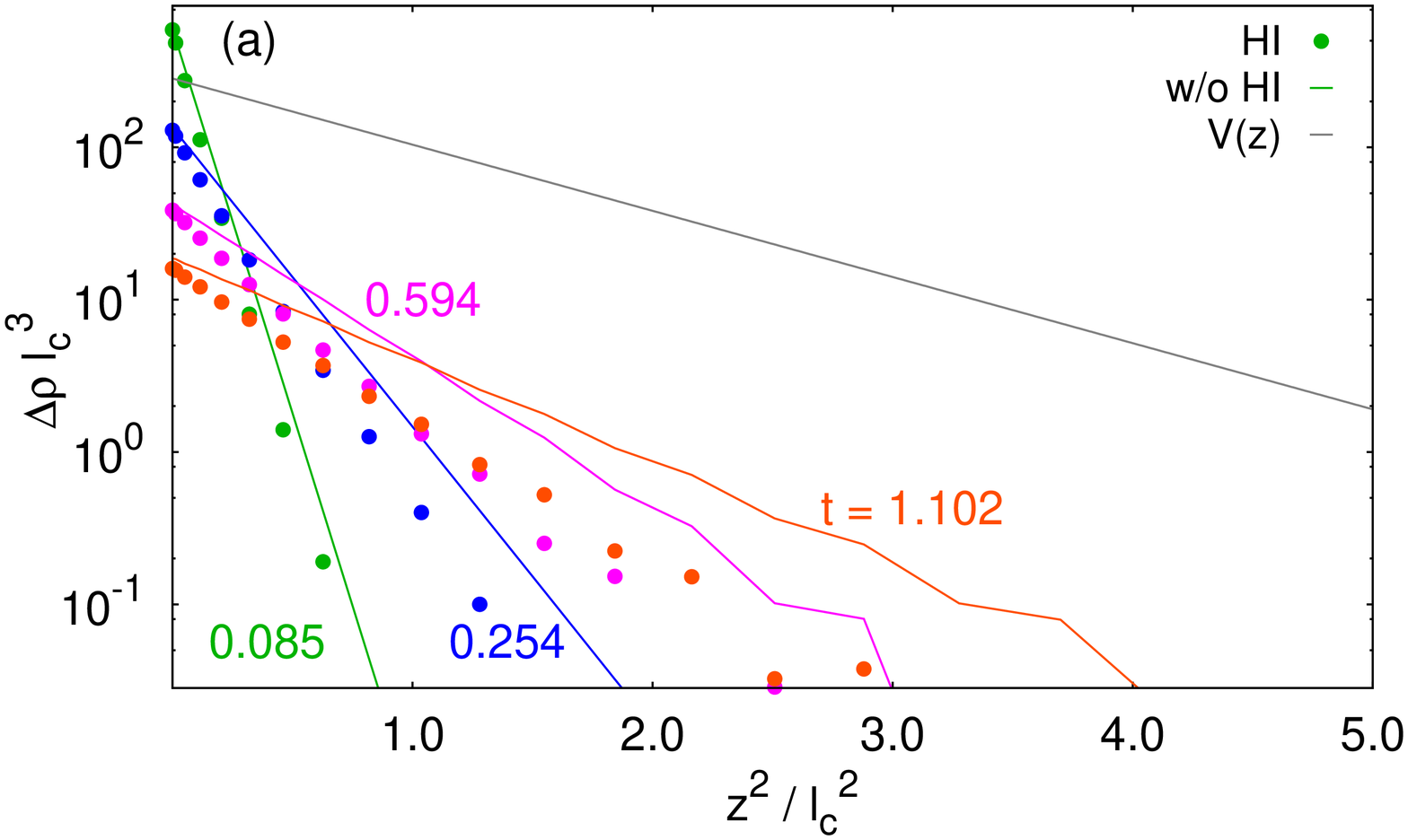,width=0.9\linewidth}
    \epsfig{file=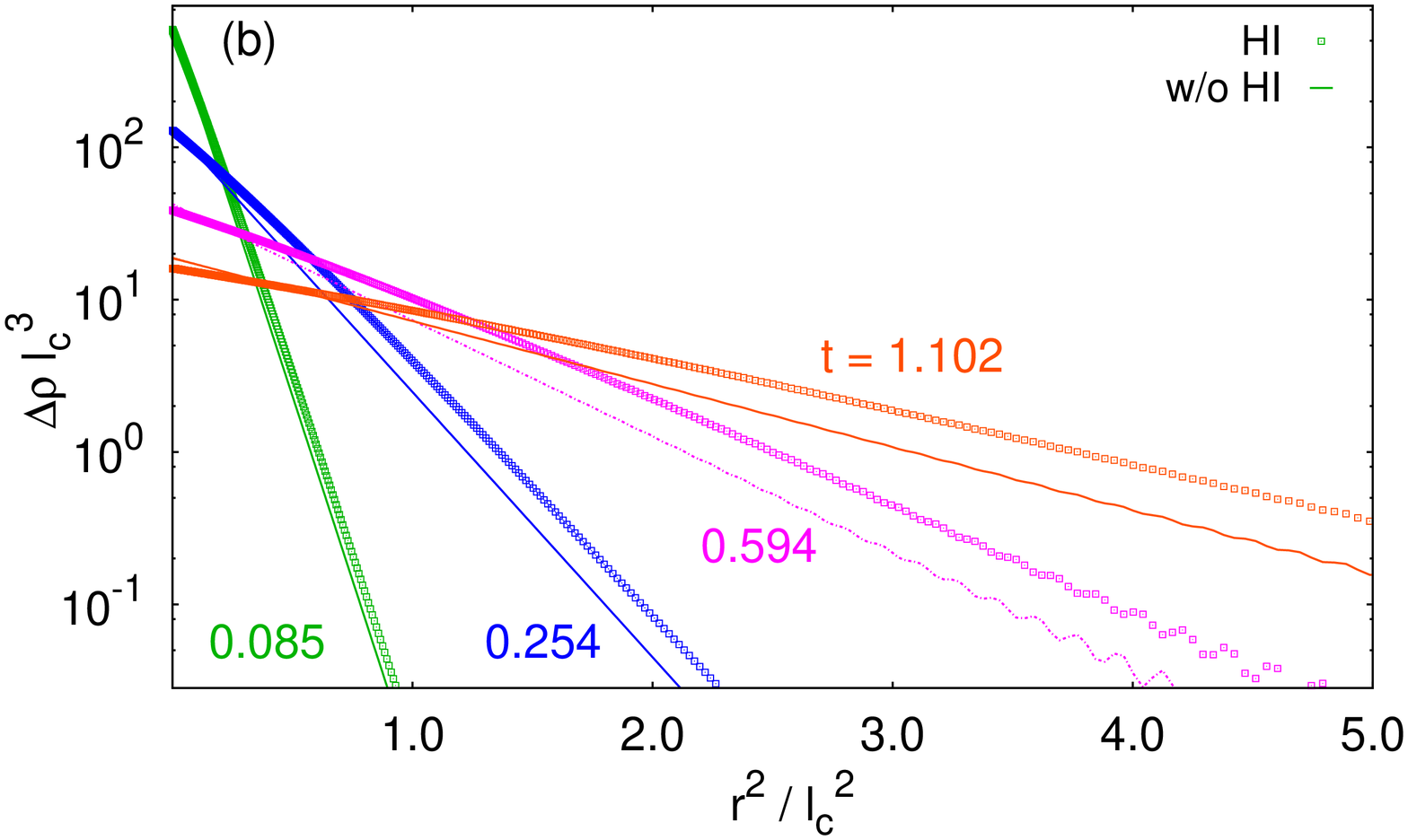,width=0.9\linewidth}
  \end{center}
  \caption{Diffusion of an initial Gaussian overdensity with $l_{\rm
      G}/\confell = 1/10$ with and without hydrodynamics. The
    horizontal axis corresponds to the squared distance from the
    origin, the vertical axis corresponds to the logarithm of the
    dimensionless overdensity profile $\confell^3 \Delta \rho$. 
    A Gaussian profile
    corresponds to a straight line.  (a) Profile for $r=0$ in
    $z$--direction at four different times. The asymptotic
    equilibrium profile~(\ref{eq:Boltzmann}) 
    is shown by the black line. (b) Profile for $z=0$ in
    $r$--direction at the same four times.  }
  \label{fig:decay_blob}
\end{figure}
Fig.~\ref{fig:decay_blob}(a) shows the time evolution of the overdensity
peak in $z$--direction, i.e., 
$\Delta\rho(0,z,t)$, and Fig.~\ref{fig:decay_blob}(b) shows the evolution in
$r$--direction, 
i.e., $\Delta\rho(r,0,t)$.  
The initially isotropic Gaussian profile roughly stays Gaussian also at later
times but becomes anisotropic. 
Even though the initial peak is not affected by the confinement, we observe
that in the presence of hydrodynamic interactions the 
peak diffuses faster in lateral $r$--direction than without them (Brownian
case), but
slower in vertical $z$--direction.  
This happens already at small times, when the system is still far from being
equilibrated in $z$--direction.  
It can be understood through particle number conservation and the 3D
incompressibility constraint ($\nabla_\bx\cdot\bu=0$) 
that a faster diffusion in $r$--direction must be accompanied with a
slower diffusion in $z$--direction: according to \eq{eq:uambient}, the
vertically directed confinement force induces a compressing flow in the
$z$--direction and, consequently, an expanding in--plane flow in the
$r$--direction. 

To elucidate the origin of the faster $r$--diffusion of the blob, we
compared the solution to the diffusion of the same peak but {\em without}
background density. 
Interestingly, the $r$--diffusion is the same for length scales
$r/\confell \lesssim 1$ (i.e., the background density does not
influence it). Only for $r/\confell \gg 1$ there is a qualitative
difference: we observe the anomalous tail in the spatial profile during the
decay of 
the peak on top of the finite background, whereas it is absent in the
decay of the peak with no background. As a conclusion, the moderate
discrepancy between lateral and vertical
diffusion of the peak (an increase of
anisotropy) is a hydrodynamic effect in regime I, occurring on scales smaller
than the width of confinement. 

\section{Summary and Conclusions}
\label{sect:sum}

We have investigated the effect of the hydrodynamic interactions on the
collective diffusion of a dilute colloidal suspension  
when the particles (having a finite hydrodynamic radius) are confined in the
vertical direction by a potential of
width $\confell$ in order to build a quasi--monolayer. Hydrodynamic
interactions have been approximated by the far--field limit at the two--body
level,  
appropriate for dilute suspensions.  The diffusion equation becomes a
nonlinear, integro--differential equation in this case.  We have
investigated collective diffusion using (i) linearized theory, (ii)
numerical solutions of the diffusion equation and (iii) truncated
Stokesian Dynamics simulations. The analysis of the linearized
  theory allows the identification of three regimes: On scales much
larger than the width of confinement $\confell$ and the
characteristic length $\lhyd$, the collective diffusion in the
  monolayer is always anomalous. For scales below $\lhyd$, the density
  evolution follows 2D normal diffusion, and for scales below
  $\confell$, the 3D normal diffusion is recovered.
We have confirmed by numerical solutions and simulations that indeed at
lateral distances $r$ much  
larger than the 
width $\confell$,
the spatial decay of density fluctuations shows instantaneously the signature
($\propto r^{-3}$) of anomalous diffusion 
characteristic for a perfect 2D monolayer.
The numerical approach also allowed the investigation of the
  transition from 3D to 2D diffusion: at small distances $r\sim
  \confell$ one already observes how the hydrodynamic interactions
  induce
faster diffusion in lateral
direction but slower in vertical direction.
Other peculiar effects also induced by the hydrodynamic
  interactions, such as the generation of regions of noticeable
  particle depletion, were observed.

  In a recent publication, Panzuela et al.~\cite{PPD17} address
  precisely the same problem of the 3D $\to$ 2D crossover in the
  diffusive dynamics of a monolayer. The numerical simulations
  presented in Ref.~\cite{PPD17} are used to measured the in--plane
  intermediate scattering function of the equilibrium density
  fluctuations, rather than the decay of macroscopic density
  profiles, as we have done in this work. Nevertheless, the
  conclusions agree in both works, which thus represent
  complementary numerical confirmations of the phenomenology
  associated to anomalous diffusion in monolayers. For completeness,
  the detailed relationship between the approach in this work and
  that in Ref.~\cite{PPD17} is discussed in App.~\ref{app:PPD}.

  Our results have far--reaching consequences for the collective
  diffusion behavior of confined systems in an infinite (or
  half--infinite medium). These systems encompass bulk colloidal
  suspensions in an external, sheet--like potential, or colloids and
  surfactants at fluid interfaces. Whenever lateral distances larger
  than the confinement width are considered, the collective diffusion
  must be considered anomalously fast. Experiments on colloidal
  monolayers indicate a hydrodynamic--induced
  enhancement of collective diffusion \cite{Zahn:1997}. A clear experimental
  signal for the $1/k$ divergence of the collective diffusion coefficient 
  can be found in Ref.~\cite{Lin:2014} which is the only experimental
  work we are aware of. The
    analysis presented in this work is intended to provide the
    theoretical framework for the analysis of the 3D $\to$ 2D
    crossover in possible future experimental realizations of the
    quasi--monolayer configuration.

  It is to be noted that, in many experimental realizations, the
  relevant configuration is a monolayer in a curved interface, e.g.,
  that of a fluid droplet. Therefore, the extension of the analysis
  to this case is desirable, with the goal of addressing the effect
  of curvature on the anomalous diffusion phenomenology. It may be
  expected that the scenario discussed here is recovered on length
  scales much smaller than the typical radius of curvature; the
  general problem is more involved and requires a detailed study.

\begin{acknowledgments}
\label{acknowl}
J.B.~thanks the German Research Foundation (DFG) for the financial
support through the Project BL 1286/2-1. A.D.~acknowledges support by
the Spanish Government through Grant FIS2014-53808-P (partially
financed by FEDER funds).

\end{acknowledgments}

\appendix

\section{Linearized equations}
\label{app:theory}

Equation~(\ref{eq:rho}) can be rewritten as
\begin{subequations}
  \begin{displaymath}
    \frac{\partial\rho}{\partial t} = - \nabla_\bx\cdot\bj ,
    \quad
    \bj := - D_0 \nabla_\bx \rho - \rho \left( \bu - \Gamma \nabla_\bx V \right) .
  \end{displaymath}    
\end{subequations}
The equilibrium solution given by \eq{eq:Boltzmann} implies
$\bu_\mathrm{eq}=0$ (after integrating by parts in \eq{eq:uambient})
and
\begin{equation}
  \label{eq:jeq}
  \bj_\mathrm{eq} = - D_0 \nabla_\bx \varrho_\mathrm{eq} 
  - \Gamma \varrho_\mathrm{eq} \nabla_\bx V = 0 .
\end{equation}
Therefore, the substitution of \eq{eq:rhopert} gives
\begin{equation}
  \bj = \varrho_\mathrm{eq} 
  \left[ - D_0 \nabla_\bx \vare + (1 + \vare) \bu \right] ,
\end{equation}
which leads to the following dynamical equation for $\vare$ after
using \eq{eq:jeq} and the incompressibility constraint
$\nabla_\bx\cdot\bu=0$:
\begin{displaymath}
  \frac{\partial \vare}{\partial t} = D_0 \nabla_\bx^2\vare - \bu\cdot\nabla_\bx\vare
  - \frac{\Gamma}{D_0} (\nabla_\bx V)\cdot\left[ D_0 \nabla_\bx\vare - (1+\vare)\bu \right] .
\end{displaymath}
The linearization of this equation around the unperturbed solution
$\vare=0$ follows easily when accounting for the fact that $\bu$ is
already of linear order in $\vare$ because $\bu_\mathrm{eq}=0$. In
this manner, \eq{eq:lineps} is obtained, while \eq{eq:linu} is simply
the already linear \eq{eq:uambient}. In particular, for the harmonic
confining potential~(\ref{eq:harmonic_pot}) the linearized equation
for $\vare$ takes the form
\begin{equation}
  \label{eq:lindiff2}
  \frac{\partial \vare}{\partial t} 
  = D_0 \left[ \nabla^2 + \frac{\partial^2}{\partial z^2} \right] \vare 
  - \frac{2 z}{\confell^2} \left[ D_0 \frac{\partial \vare}{\partial z}
    - \be_z\cdot\bu \right] .
\end{equation}
The expansion~(\ref{eq:Hermite}) can be inverted as
\begin{multline}
  c_n(\bk,t) = \int d^2\br \, \mathrm{e}^{-i\bk\cdot\br} 
  \, \int_{-\infty}^{+\infty} \frac{dz}{\confell}\; 
  \frac{\mathrm{e}^{-(z/\confell)^2}}{\sqrt{\pi} \, n! \, 2^n} \\ 
  \times H_n\left(\frac{z}{\confell}\right) \, \vare(\br,z,t) ,
\end{multline}
and \eq{eq:c} is obtained from \eq{eq:lindiff2} by using that the
Hermite polynomials satisfy 
\begin{displaymath}
  \frac{d^2 H_n(\zeta)}{d \zeta^2} - 2 \zeta \frac{d H_n(\zeta)}{d \zeta} + 2 n H_n(\zeta) 
  = 0.
\end{displaymath}
The function $\psi_n(\bk, t)$ appearing in \eq{eq:c} is given as
\begin{align}
  \label{eq:psi2}
  & \psi_n(\bk,t) := \frac{\confell \lhyd\,k}{D_0}
  \int d^2\br \; \mathrm{e}^{-i\bk\cdot\br} 
  \\ 
  &\times \int_{-\infty}^{+\infty} \frac{dz}{\confell} \; 
  \frac{2(z/\confell) \mathrm{e}^{-(z/\confell)^2}}{\sqrt{\pi} \, n! \, 2^n} \; 
  H_n \left(\frac{z}{\confell}\right) \; \be_z\cdot\bu(\br,z,t) ,
  \nonumber
\end{align}
which is computed in App.~\ref{app:psi}.

\section{Calculation of the coefficients $\Omega_{nm}$}
\label{app:psi}

In order to compute $\psi_n$ defined by \eq{eq:psi2}, one first
calculates $\be_z\cdot\bu$ given by \eq{eq:linu} with the harmonic
potential~(\ref{eq:harmonic_pot}) (for the purpose of this Appendix,
the explicit time dependence will be dropped from the notation):
\begin{multline}
  \be_z\cdot\bu(\br,z) = - \frac{2 D_0 \confell \rho_0^\mathrm{(2D)}}{\sqrt{\pi}}
  \int\frac{d^2\br'}{\confell^2} \int_{-\infty}^{+\infty} \frac{dz'}{\confell} \;
  \frac{z'}{\confell} \\ 
  \times \mathrm{e}^{-(z'/\confell)^2}  \nonumber 
   \vare(\br',z') \; \be_z\be_z : \bom (\br-\br'+\be_z(z-z')).
\end{multline}
This expression can be evaluated by inserting the 3D Fourier transform
of the Oseen tensor \cite{KiKa91},
\begin{multline}
  \bom(\br+\be_z z) = 3\pi\sigh \int \frac{d^2\bk}{(2\pi)^2} 
  \int_{-\infty}^{+\infty} \frac{dk_z}{2\pi} \; 
  \frac{\mathrm{e}^{i\bk\cdot\br + i k_z z}}{k^2+k_z^2} 
  \\ 
  \times \left[ \mathcal{I} - \frac{(\bk+\be_z k_z)(\bk+\be_z
      k_z)}{k^2+k_z^2} \right] ,
  \nonumber
\end{multline}
and using that
\begin{displaymath}
  \int_{-\infty}^{+\infty} \frac{d k_z}{2\pi} \; 
  \frac{\mathrm{e}^{i k_z (z-z')}}{(k^2+k_z^2)^2}
  = \frac{1}{4 k^3} \mathrm{e}^{-k |z-z'|} \left( 1 + k |z-z'| \right) ,
\end{displaymath}
so that the integrals over $k_z$ and $\br'$ can be carried out,
which results in
\begin{widetext}
\begin{displaymath}
  \be_z\cdot\bu(\br,z) = - \frac{2 D_0}{\sqrt{\pi} \confell \lhyd}
  \int \frac{d^2\bk}{(2\pi)^2} \frac{\mathrm{e}^{i\bk\cdot\br}}{k} \;
  \sum_{m=0}^\infty c_m(\bk) 
  \int_{-\infty}^{+\infty} \frac{dz'}{\confell} \;
  \frac{z'}{\confell} \mathrm{e}^{-(z'/\confell)^2}
  H_m\left(\frac{z'}{\confell}\right)
  \mathrm{e}^{-k |z-z'|} \left( 1 + k |z-z'| \right) ,\nonumber
\end{displaymath}
after using the definition~(\ref{eq:Lhydro_gen}) and the
expansion~(\ref{eq:Hermite}). Therefore, when this expression is
inserted in \eq{eq:psi2}, one obtains \eq{eq:psi} with the
coefficients
\begin{equation}
  \label{eq:Omega}
  \Omega_{nm}(q) := 4 \int_{-\infty}^{+\infty}d\zeta \; \int_{-\infty}^{+\infty}d\zeta' \;
  \zeta \zeta' \mathrm{e}^{-(\zeta^2+\zeta'^2)} 
  H_n(\zeta) H_m(\zeta') \, 
  \mathrm{e}^{-q |\zeta-\zeta'|} \left( 1 + q |\zeta-\zeta'|  \right),
\end{equation}
\end{widetext}
in terms of the dimensionless quantities
\begin{displaymath}
  \zeta := \frac{z}{\confell} ,
  \qquad
  q := \confell k .
\end{displaymath}
It is manifest that $\Omega_{nm}(q)$ is symmetric in the indices.
Furthermore, a change of variables $\zeta\to -\zeta$, $\zeta'\to
-\zeta'$ in the integrals shows that $\Omega_{nm}(q)$ vanishes if $n$
and $m$ have different parity. It is possible to simplify
\eq{eq:Omega} and eventually express it in terms of the error
function. We are mainly interested, however, in the asymptotic
behaviors in the limits $q\gg 1$ and $q\ll 1$, and this can be derived
directly from \eq{eq:Omega}.

When $q\to \infty$, one can evaluate \eq{eq:Omega} using Laplace's
formula \cite{BeOr78}, because the integral is dominated by the value
of the integrand near $\zeta-\zeta'=0$. Introducing the new variables
$\mu=\zeta+\zeta'$, $\sigma=\zeta-\zeta'$, one can write
\begin{multline}
  \Omega_{nm}(q) = \frac{1}{2} \int_{-\infty}^{+\infty} d\sigma\;
  \mathrm{e}^{-q|\sigma|} \left( 1 + q |\sigma| \right)
  \mathrm{e}^{-\sigma^2/2}
  \\
  \times \int_{-\infty}^{+\infty} d\mu\; \mathrm{e}^{-\mu^2/2}
  (\mu^2-\sigma^2) H_n\left(\frac{\mu+\sigma}{2}\right)
  H_m\left(\frac{\mu-\sigma}{2}\right)
  \nonumber
\end{multline}
As $q\to\infty$, this expression can be approximated as
\begin{multline}
  \Omega_{nm}(q) \sim \frac{1}{2} \int_{-\infty}^{+\infty} d\sigma\;
  \mathrm{e}^{-q|\sigma|} \left( 1 + q |\sigma| \right)
  \\
  \times \int_{-\infty}^{+\infty} d\mu\; \mathrm{e}^{-\mu^2/2}
  \mu^2 H_n\left(\frac{\mu}{2}\right)
  H_m\left(\frac{\mu}{2}\right)
  \nonumber
\end{multline}
from where one concludes that $\Omega_{nm}(q)\sim 1/q$.

In the opposite limit $q\to 0$, one can Taylor--expand the integrand
in \eq{eq:Omega} because $|\mathrm{e}^{- q s} \left( 1 + q s \right)|
\leq 1$ for $s\geq 0$, and thus the integral converges uniformly in
$q$. One has
\begin{displaymath}
  \mathrm{e}^{- q|\zeta-\zeta'|} \left( 1 + q |\zeta-\zeta'|
  \right) = 1 - \frac{1}{2} q^2 (\zeta^2+\zeta'^2-2\zeta\zeta') 
  + o(q^3) .
\end{displaymath}
When this expression is substituted in \eq{eq:Omega}, the two
integrals factorize. They can be computed explicitly by expressing the
powers of $\zeta$ and $\zeta'$ in terms of the Hermite polynomials and
using the associated orthonormality relations:
\begin{displaymath}
  \int_{-\infty}^{+\infty} ds\; \mathrm{e}^{-s^2} H_a(s) H_b(s) 
  = \sqrt{\pi}\, a! \, 2^a \, \delta_{a,b} .
\end{displaymath}
For the particular case that both indices $n$, $m$ are even, one
obtains
\begin{subequations}
  \label{eq:Omegasmallk}
  \begin{equation}
    \Omega_{00} \sim \pi q^2 - \sqrt{8\pi}\, q^3 + o(q^4) ,
  \end{equation}
  \begin{equation}
    \Omega_{20} \sim 4\pi q^2 + o(q^3) , 
  \end{equation}
  \begin{equation}
    \Omega_{22} \sim 16\pi q^2 + o(q^3) , 
  \end{equation}
  \begin{equation}
    \Omega_{nm} \sim o(q^3) \quad \mathrm{if}\; n\geq 4 \; \mathrm{or}\; m\geq 4 .
  \end{equation}
\end{subequations}

\section{Adiabatic elimination of the fast modes}
\label{app:adiabatic}

We introduce the short--hand notations $q:= \confell k$, $\lambda :=
\lhyd k$ and $\tau := D_0 t/\confell^2$, and define the
infinite--dimensional column vector $\mathbf{c} := (c_2 \; c_4 \;
\dots)^\dagger$, so that the dynamical equations~(\ref{eq:fullc}) for
$n\geq 2$ can be rewritten in compact form as\footnote{To avoid a
  cumbersome notation, we ignore the fact that $n$ and $m$ represent
  even numbers, but the indices of the components of vectors and
  matrices must be natural numbers. This should not create ambiguity
  because the simplicity of the expressions is self-explanatory.}
\begin{displaymath}
  \pi \, n! \, 2^n \, \lambda \frac{d c_n}{d \tau} 
  = - \left( \mathcal{B}\cdot\mathbf{c} + c_0 \mathbf{s} \right)_n 
  \qquad 
  (n\geq 2) ,
\end{displaymath}
in terms of the symmetric matrix
\begin{displaymath}
  \mathcal{B} := \mathrm{diag}\left[ \pi \, n! \, 2^n \, \lambda \, (q^2 + 2 n) \right]
  + (\Omega_{mn}) ,
\end{displaymath}
and the column vector $\mathbf{s} := ( \Omega_{20} \; \Omega_{40} \;
\dots )^\dagger$. The ``adiabatic enslaving'' of these fast modes
(notice that $\mathcal{B}$ 
does not vanish as $q\to 0$ provided $\lambda\neq 0$)
gives the relationship
\begin{displaymath}
  \frac{d \mathbf{c}}{d \tau} = 0 
  \quad\Rightarrow\quad
  \mathbf{c}_\mathrm{enslaved} = - c_0 \, \mathcal{B}^{-1}\cdot\mathbf{s} .
\end{displaymath}
Inserting it into the dynamical equation for the slow mode $c_0$
(\eqs{eq:fullc}) yields
\begin{displaymath}
  \frac{d c_0}{d \tau} \approx - \left[ q^2 + \frac{\Omega_{00}(q)}{\pi \lambda} \right] c_0
  - \frac{1}{\pi \lambda} \mathbf{s}\cdot\mathbf{c}_\mathrm{enslaved} ,
\end{displaymath}
which becomes the linear \eq{eq:dync0} with the diffusion coefficient
\begin{equation}
  \label{eq:longD2}
  \frac{D(k)}{D_0} - 1 \approx \frac{1}{\pi \lambda q^2} \left[ \Omega_{00}(q) - 
    \mathbf{s}\cdot\mathcal{B}^{-1}\cdot\mathbf{s} \right] .
\end{equation}
For consistency with the assumption of ``adiabatic enslaving'', this
expression is meaningful only in the limit $q\to 0$. From
\eqs{eq:Omegasmallk} and
\begin{displaymath}
  \mathbf{s} = \left(
    \begin{array}[c]{c}
      4\pi q^2 + o (q^3) 
      \\
      o (q^3) 
      \\
      \vdots
    \end{array}
    \right) ,
  \end{displaymath}
  \begin{displaymath}
    \mathcal{B}^{-1} = 
    \frac{1}{2\pi\lambda} \mathrm{diag}\left(\frac{1}{n! \, 2^n \, n}\right)
    + o\left(\frac{q}{\lambda}\right)^2 ,
\end{displaymath}
one gets for \eq{eq:longD2} the expression
\begin{displaymath}
  \frac{D(k)}{D_0} - 1 \approx \frac{1}{\lambda} - \sqrt{\frac{8}{\pi}} \frac{q}{\lambda}
  - \frac{q^2}{2\lambda^2} + o\left(\frac{q^2}{\lambda}, \frac{q^6}{\lambda^2} \right) ,
\end{displaymath}
which is \eq{eq:longD}. The criterion for not retaining higher order
terms in this expansion is that they lead to positive powers of $k$ in
\eq{eq:longD}. This ultimate goal is also the reason for the careful
bookkeeping in powers of both $q$ \emph{and} $\lambda$ when deriving
the expansion.

\section{Comparison with Ref.~\cite{PPD17}}
\label{app:PPD}

Panzuela et al.~\cite{PPD17} obtain a theoretical result for the
  short--time collective diffusion coefficient 
$D^\mathrm{(short)}(k)$ by studying
the decay of in--plane equilibrium
density fluctuations, i.e., the intermediate scattering function
\begin{eqnarray}
  \label{eq:scatteringF}
  F(\bk, t) 
  & = &
  \langle \hat{\rho}(\bk, t) \hat{\rho}^*(\bk, 0) \rangle 
\end{eqnarray}
where $\langle\dots\rangle$ denotes the average over the equilibrium
distribution in the initial state, and $\hat{\rho}$ is the microscopic
density field. The theoretical analysis assumes a dilute system
  and focuses onto the short--time regime, i.e., times $t\to 0$ so
that the colloidal particles are displaced by an amount much
smaller than the mean interparticle separation. In such case, one
  assumes
  \begin{displaymath}
    F(\bk, t) = F(\bk,0) \exp( - k^2 D^\mathrm{(short)}(k) t)
    \qquad (t\to 0),
  \end{displaymath}
  with the definition 
\begin{displaymath}
  D^\mathrm{(short)}(k) := - \frac{1}{k^2} \left[ \frac{1}{F(\bk,t)}
    \frac{\partial F(\bk,t)}{\partial t} \right]_{t=0} .
\end{displaymath}
The theoretical model we have devised concerns the time evolution of the
average density 
$\rho = \langle\hat{\rho}\rangle$, see \eq{eq:density}. Nevertheless,
the same result can be obtained for the short--time dynamics starting
with our linearized equations for the time evolution of a density
  fluctuation (\eqs{eq:lindiff}).  
In terms of the projected average 2D density (see
\eqs{eq:inplanerho}),
\begin{equation*}
  \rho^\mathrm{(2D)}(\bk, t) = \rho^\mathrm{(2D)}_0 c_0 (\bk, t)
  \qquad
  (\bk\neq 0) ,
\end{equation*}
one can define the short--time diffusion coefficient as
\begin{displaymath}
  D^\mathrm{(short)}(k) := - \frac{1}{k^2} \left[ \frac{1}{\rho^\mathrm{(2D)}(\bk,t)}
  \frac{\partial \rho^\mathrm{(2D)}(\bk,t)}{\partial t} \right]_{t=0} .
\end{displaymath}
Use of \eqs{eq:fullc} gives the expression
\begin{equation*}
  \frac{D^\mathrm{(short)}(k)}{D_0} -1 = 
    \frac{1}{\pi \lhyd k} 
    \sum_{m=0}^{\infty} \frac{\Omega_{0m}(\confell k)}{(\confell k)^2} 
    \frac{c_m(\bk,0)}{c_0(\bk,0)} .
\end{equation*}
This is not a well--defined system--characteristic quantity due to the
dependence on the specific initial conditions $c_m(\bk,0)$. However,
one can restrict consideration to initial perturbations with $c_m=c_0
\delta_{m,0}$, as is actually done in \eq{eq:scatteringF} when
performing the average over equilibrium configurations, for which
the different modes $c_m$ are uncorrelated: in such case, after
evaluating $\Omega_{00}$, see App.~\ref{app:psi}, one obtains
\begin{widetext}
\begin{equation}
  \label{eq:shortD}
  \frac{D^\mathrm{(short)}(k)}{D_0} - 1 
  = \frac{1}{\pi \lhyd k} \frac{\Omega_{00}(\confell k)}{(\confell k)^2} 
  = \frac{1}{\lhyd k} \left\{
    \left[ 1 + (\confell k)^2 \right] \mathrm{e}^{\frac{(\confell k)^2}{2}}
    \mathrm{erfc}\left( \frac{\confell k}{\sqrt{2}}\right) 
    - \sqrt{\frac{2}{\pi}} \; \confell k
  \right\} ,
\end{equation}
\end{widetext}
in terms of the complementary error function $\mathrm{erfc}(q)$. This
result coincides exactly with Ref.~\cite[{}][{Eq.~(18)}]{PPD17}, with
the notation $\confell = \sqrt{2}\, \delta$
\cite[{}][{Eq.~(1)}]{PPD17} and $\lhyd = 2 a/3 \phi$
\cite[{}][{Eq.~(20)}]{PPD17}. It is to be compared with
\eq{eq:longD2} derived in the opposite, long--time limit. Both
coefficients agree on the dominant, anomalous--diffusion behavior at
large scales. Expression~(\ref{eq:shortD}) is not restricted to the
small $q$ limit, but at the price of choosing a certain set of initial
conditions \textit{ad hoc}. Equation~(\ref{eq:longD2}), on the
contrary, is valid only in the limit $q\to 0$, but it incorporates
naturally the irrelevance of the initial conditions through the
``adiabatic enslaving''.




\end{document}